\documentclass[aps,prl,twocolumn,10pt,showpacs,superscriptaddress]{revtex4}

\usepackage{graphicx}
\usepackage{amsmath,units}
\usepackage{mathrsfs}
\usepackage{color} 

\begin{document}
\title{Optical trapping below the diffraction limit with a tunable beam waist using super-oscillating beams}
\author{Harel Nagar}
\affiliation{Raymond \& Beverly Sackler School of Chemistry, Tel Aviv
	University, Tel Aviv 6997801, Israel}
\author{Tamir Admon}
\affiliation{Raymond \& Beverly Sackler School of Chemistry, Tel Aviv
	University, Tel Aviv 6997801, Israel}
\author{Doron Goldman}
\affiliation{he Future Scientists Center- Alpha Program at Tel Aviv Youth University}
\author{Amir Eyal}
\affiliation{he Future Scientists Center- Alpha Program at Tel Aviv Youth University}

\author{Yael Roichman}
\email{roichman@tauex.tau.ac.il}
\affiliation{Raymond \& Beverly Sackler School of Chemistry, Tel Aviv
	University, Tel Aviv 6997801, Israel}

\begin{abstract}
Super-oscillating beams can be used to create light spots whose size is below the diffraction limit with a side ring of high intensity adjacent to them. Optical traps made of the super-oscillating part of such beams exhibit superior localization of submicron beads compared to regular optical traps. Here we focus on the effect of the ratio of particle size to trap size on the localization and stiffness of optical traps made of super-oscillating beams.  We find a non-monotonic dependence of trapping stiffness on the ratio of particle size to beam size.  Optimal trapping is achieved when the particle is larger than the beam waist of the super-oscillating feature but small enough not to overlap with the side ring.    
\end{abstract}

\maketitle

In the early 70s, Artur Ashkin showed that a weakly focused laser beam can draw small particles with high refractive index towards its center and move them in the direction of light propagation \cite{Ashkin1970}. A major breakthrough in this field happened in 1986 when Ashkin demonstrated the single beam optical gradient force traps \cite{Ashkin1986}, known nowadays as optical tweezers. Since then, optical trapping application has become a powerful tool used in physics and biology. However, the size of an optical trap is limited by the smallest spot which collimated light can be focused to using an annular aperture, as discussed in 1873 by Ernst Abbe \cite{Abbe1873} and later by Lord Rayleigh \cite{Rayleigh1879}.  The diffraction limit of light determining this minimal beam size is given by $w=0.38\lambda/$NA, where $w$ is the beam waist defined as the full width at half maximum of the beam, $\lambda$ is the wavelength of the beam, and NA is the numerical aperture of the focusing lens. In 1952 G. Toraldo di Francia suggested theoretically that by phase modulations one can achieve optical features below the diffraction limit \cite{Francia1952}. In the 90's the concept of super oscillation (SO) was first introduce by Michel Berry for bandlimited functions that locally oscillate faster than their highest Fourier component \cite{Berry1994}. In optics, the SO phenomena was used to generate optical beams with features smaller than the diffraction limit. Over the last 20 years SO beams were generated using different methods \cite{Rogers2013,Huang2007, Eliezer2016} and applied for super-resolution imaging \cite{Zheludev2008,Zheludev2012}. 

The effect of particle size, beam waist, and wavelength on the stiffness of optical trapping was studied theoretically for different scattering regimes \cite{Harada1996,Sato2007,Dunlop2008}. Experimental verification of these predictions is challenging since neither beam size, wavelength, nor particle size can be changed continuously to provide a clean comparison \cite{Heckenberg1994,Arimondo2002,Samadi2010,Dharmadhikari2013}. Naturally, all previous measurements focused on diffraction-limited optical traps. Previously, we observed that a significant enhancement of optical trapping strength and localization occurred when a 490 nm particle was trapped in the SO part of a SO beam \cite{SinghSO2017}. Here we study this effect in more detail. We use the unique feature of SO beams, namely, the ability to change continuously the beam waist and to focus the beam to below the diffraction limit, to measure the effect of particle size and beam waist on the stiffness of SO beam based optical traps. We repeat our experiments for four different particle sizes changing the beam waist in real time.

\begin{figure}[htbp]
	\centering
	\includegraphics[width=0.8\linewidth]{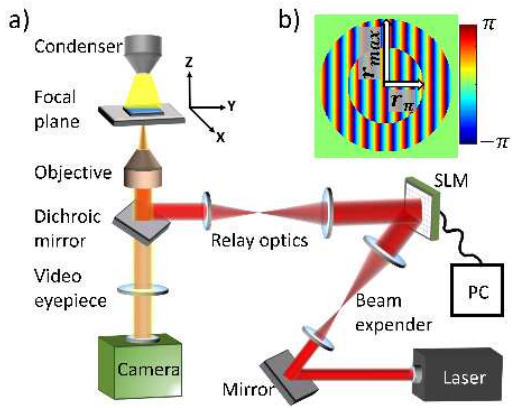}
	\caption{ (a) Schematic diagram of the holographic optical tweezer system. An infrared laser beam is projected and expanded to cover the face of a SLM. The hologram encoded on the beam by the SLM is formed in the sample plane of the microscope. (b) Phase mask of a SO beam.
	}
	\label{fig:fig1}
\end{figure}

To generate the SO beams we project an infrared laser beam (1083 nm, Keopsys, KPS-KILAS-TRAPP1083-20-PM-CO) on a spatial light modulator (SLM) encoding a binary phase mask (see Refs.\cite{Remez2015,Pedro2004}). The wave function encoded by the SLM is  (see Fig.~\ref{fig:fig1}b):
\[ \psi(r,x) = \left\{ 
\begin{array}{l l}
\exp\{i(k_cx_{s}+\pi)\} & r\leq r_{\pi}\\
\exp\{i(k_cx_{s})\} & r_{\pi}\leq r\leq r_{max}
\end{array} \right.\]
where  $0< r_{\pi}< r_{max}$ , $r_{max}$ is the radius of the mask aperture at the pupil plane, $r$ is the radial coordinate, $k_c$ is the carrier wave number to separate the SO beam from the unwanted zero order beam, and $x_s$ is normalized by the number of pixels in the SLM. For simplicity we define $\tilde{r}_\pi=r_{\pi}/r_{max}$. The reflected beam from the SLM is transmitted through relay optics into an inverted X71 Olympus microscope. The focal plane of the microscope objective is conjugated to the SLM plane. Therefore, we obtain in the sample plane the Fourier transform of the reflected beam $FT\{\psi(r,x)\}$, which is a SO beam (see Fig.~\ref{fig:fig1}). By changing $\tilde{r}_\pi$ the size of the central spot of the SO beam can be controlled and reduced to below the diffraction limit. It should be noted that for smaller spot sizes the intensity of the central spot reduces dramatically.

\begin{figure}[htbp]
	\centering
	\includegraphics[width=1\linewidth]{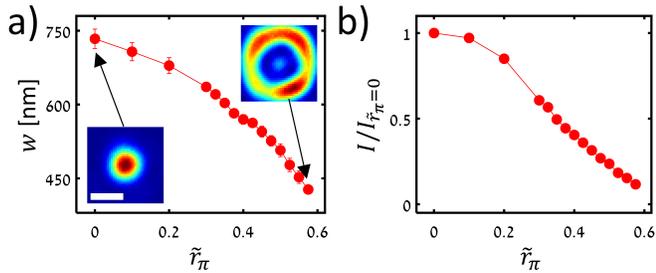}
	\caption{ a-b) The measured beam waist and peak intensity of the central spot as a function of $\tilde{r}_\pi$ are shown in (a) and (b), respectively. Inset: experimental beam intensity profile of SO beams with $\tilde{r}_\pi=0$ and $\tilde{r}_\pi=0.575$. Scale bar: 1$\mu$m.  
	}
	\label{fig:fig2}
\end{figure}

We create a large range of SO optical traps with a beam waist in the range of 427-733 nm, since the phase mask aperture is smaller than the lens aperture diameter, the effective NA (NA$_{eff}$=0.55) is smaller than the NA of the objective (NA$_{obj}$=1.4). The beam waist size can be compared to the theoretical diffraction-limit waist size of our system ($750$ nm). The structure of our SO beam at the focal plane of the microscope includes a central spot surrounded by a high intensity light ring \cite{SinghSO2017} (see inset of Fig.~\ref{fig:fig2}a). From Fig.~\ref{fig:fig2} it can be seen that beam waist size and peak intensity of the central spot reduce with an increase in $\tilde{r}_\pi$. The beam waist decreases down to 427 nm for $\tilde{r}_\pi$=0.575 with a peak intensity that is about a tenth of its intensity in a $\tilde{r}_\pi=0$ beam (Fig.~\ref{fig:fig2}b).

\begin{figure}[htbp]
	\centering
	\includegraphics[width=0.8\linewidth]{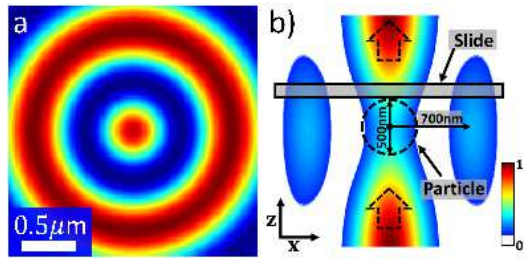}
	\caption{ (a) Simulated intensity profile of SO beam $\tilde{r}_\pi$=0.55 at the focal plane. (b) Intensity cross section the SO beam in relation to the sample positioning. The particle is trapped in the central spot at the focal plane against the glass wall of the sample chamber.
	}
	\label{fig:fig3}
\end{figure}

In each experiment we trapped a polystyrene bead in the central spot of the beam against the cover glass as depicted in Fig.~\ref{fig:fig3} ensuring it doesn't hop to the side ring. Trapping in the axial direction is not possible in this type of beams since the gradient force points away from the focal plane and not toward it like in regular optical tweezers \cite{SinghSO2017}. We then changed the beam waist while the particle remained trapped to a new value by changing $\tilde{r}_\pi$. During the whole experiment particle position was recorded using conventional video microscopy \cite{Grier1996}. The trapping quality for each $\tilde{r}_\pi$ value was characterized by calculating the trap stiffness and the localization area of the bead. We adjusted the laser power at each $\tilde{r}_\pi$ value so that the peak intensity of the central spot will be kept constant throughout the experiment to obtain a meaningful comparison. We repeated these experiments for four different particle sizes listed in Table~\ref{tab:sizes}. 

\begin{table}[htbp]
	\centering
	\caption{\bf Polystyrene particle }
	\begin{tabular}{ccc}
		\hline
		Diameter [nm] & Company & Catalog code \\
		\hline
		490 & Invitrogen &  F8812 \\
		780 & Polysciences, Inc. & 07766 \\
		1100 & Invitrogen & F8775 \\
		1370 & Bangs Laboratories, Inc. & PS04N \\
		\hline
	\end{tabular}
	\label{tab:sizes}
\end{table}
We define the particle localization $\sigma$ as the standard deviation of the particle displacement within the trap. Trap stiffness $k$ is extracted from the power spectrum of the particle displacements by fitting it to a Lorentzian \cite{Kirstine2004}:
\begin{equation}
P(f) = \frac{k_B T}{\pi^2\gamma\left[\left(\frac{k}{2\pi\gamma}\right)^2 +f^2\right]}
\label{eq:pow}
\end{equation}
Where $P(f)=\left|\tilde{x}(f)\right|^2/T_{msr}$ is the power spectrum, $\tilde{x}(f)$ is the Fourier transform of the trapped particle position, $T_{msr}$ is the duration of the  measurement, $k_B$ is Boltzmann's constant, $T$ is the absolute temperature, $\gamma=6\pi\eta R$ is the drag coefficient of the bead, $\eta$ is the suspending fluid viscosity, $R=a/2$ is the particle radius, and $a$ is the particle diameter.

We start by examining the absolute localization of particles of different sizes. We find that for $\tilde{r}_\pi$=0 localization is improved as particle size increases in accord with previous studies \cite{Heckenberg1994,Dimova2006}. As we increase $\tilde{r}_\pi$ the localization of the particles changes, however, the trend of change depends on the size of the trapped particle. We observe that trapping becomes unstable at large $\tilde{r}_\pi$. We define $r_{\pi}^{stable}$ to be the maximal value for which trapping is stable, this corresponds to the minimal stable trap size. The absolute localization value for all particle sizes for $\tilde{r}_\pi$=0 and $r_{\pi}^{stable}$ are given in Table~\ref{tab:loc}. We note that $r_{\pi}^{stable}$ depends on particle size. For the small particle $a=490$ nm we observe an enhancement of the localization with an increase in $\tilde{r}_\pi$ in agreement with a recent report \cite{SinghSO2017}. However, for larger particles this is not the case, even in conditions of stable trapping. For example, for the biggest particle (1370 nm) the localization deteriorates significantly as $\tilde{r}_\pi$ is increased.

\begin{table}[htbp]
	\centering
	\caption{\bf Localizations }
	\begin{tabular}{cccc}
		\hline
		Diameter &$\sigma(\tilde{r}_\pi$=0$)$ &   $r_{\pi}^{stable}$ & $\sigma(r_{\pi}^{stable})$ \\
		\ [nm] & [nm] &  & [nm] \\
		\hline
		490 & 78$\pm$8 & 0.550 &  56$\pm$6 \\
		780 & 42$\pm$2 & 0.500 & 43$\pm$3 \\
		1100 & 33$\pm$3 & 0.425 & 39$\pm$2 \\
		1370 & 34$\pm$3 & 0.375 & 80$\pm$10 \\
		\hline
	\end{tabular}
	\label{tab:loc}
\end{table}

\begin{figure}[h!]
	\centering
	\includegraphics[width=0.8\linewidth]{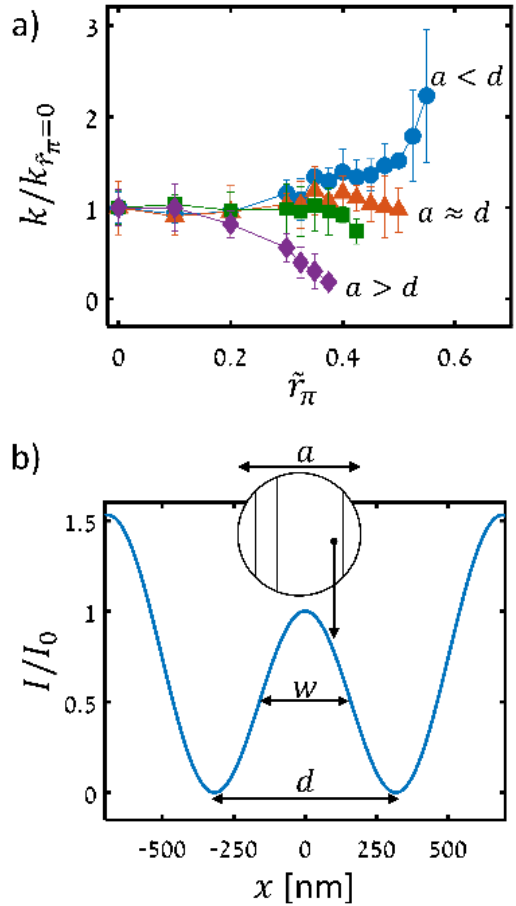}
	\caption{ a) The change in the normalized trap stiffness $k/k_{\tilde{r}_\pi=0}$ of different particle sizes as $\tilde{r}_\pi$ increases (decreasing beam waist size). The diameters of the shown particles are 490 nm (blue circles), 780 nm (orange triangles), 1100 nm (green squares), and 1370 nm (purples diamonds). b) An illustration comparing particle diameter $a$, beam waist $w$ and distance to the minima of the SO beams $d$ to the calculated intensity cross-section of an SO beams with  $\tilde{r}_\pi=0.55$.
	}
	\label{fig:fig4}
\end{figure}

We now quantify the relative change in trap stiffness as a function of $\tilde{r}_\pi$ (Fig.~\ref{fig:fig4}a). We define the distance of the intensity minima of the SO beam from its center as $d$ as shown in Fig.~\ref{fig:fig4}b where an intensity cross section of the SO beam is plotted.  In our experiments $d$ varied between 640 to 810 nm. Interestingly, we find that for particles whose diameter is smaller than $d$ decreasing the beam waist of the central spot of the SO beam improves localization and increases trap stiffness, for particles of $a\approx d$ trap stiffness and localization are relatively unaffected by the change in the SO beam structure and for particles with $a>d$ trapping quality deteriorates with the decrease in the central spot size.    

\begin{figure}[h!]
	\centering
	\includegraphics[width=1\linewidth]{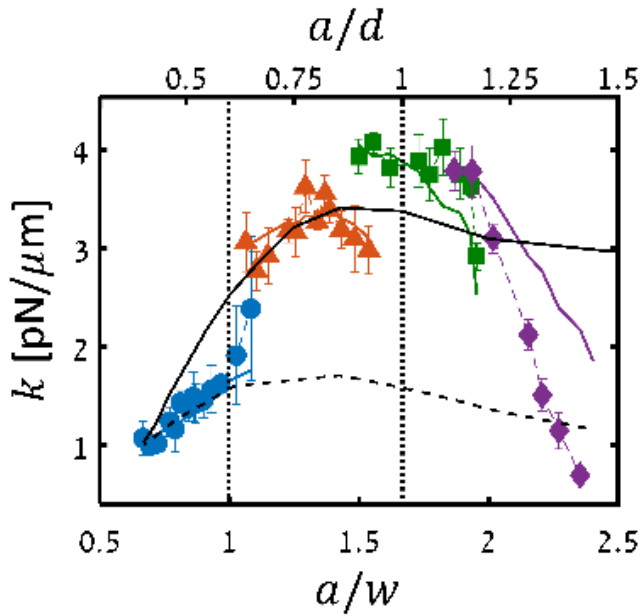}
	\caption{Effect of particle size on trapping stiffness in experiment (symbols) and simulations (solid colored lines) compared to $w$ and $d$. The diameters of the shown particles are 490 nm (blue circles), 780 nm (orange triangles), 1100 nm (green squares) and 1370 nm (purples diamonds), modified Rayleigh scattering simulation results are presented as solid lines in the same color as the experiment results. The black dash and solid lines are Gaussian beam trapping simulations performed using the modified Rayleigh scattering force calculation and the Mie scattering force calculation, respectively (see text for details). 
	}
	\label{fig:fig5}
\end{figure}

In Fig.~\ref{fig:fig5} we plot the trap stiffness as a function of the ratio between particle diameter and the beam waist size. We find that the combination of experiments results in single non-monotonous curve. Clearly, as long as $a<d$ the trap stiffness, for equal peak beam intensity, increases even for $a \ge w$. This happens since the cross-section of the particle and the light increases as does the trapping force, and the particle is too small to be affected by the side ring. For the intermediate regime $(a/w>1$ and $a/d<1)$ no significant change in the trapping stiffness was observed. However, once $a>d$ trapping stability is compromised, since in this range of particle size the overlap between the side ring and the particle is inescapable. 

We suggest that the destabilization of particles in SO beams is a result of the interaction of the trapped particles with the high intensity side ring of the beam. This interaction is enhance as $\tilde{r}_\pi$ increases due to the increase of the relative power of the side ring. 
To support our hypothesis we performed a set of numerical calculations. In order to calculate the effect of particle size on the trap stiffness we perform a two step calculation, First we calculate the optical forces acting on the trapped particle due to an SO beam and then use these forces to simulate particle motion within the traps. 
To estimate the optical forces, we consider the particle to be composed of 10 nm wide slices (see Fig.~\ref{fig:fig4}b). We calculate the gradient force (in the Rayleigh regime  ($R\ll\lambda$) \cite{Dholakia2008,Padgett2002} ) on each slice according to $F_{slice}=\frac{\alpha}{2}\nabla I$, where $\alpha$ is the particle polarizability \cite{Ashkin1986}. The total force on the bead ($F_{bead}$) is given by integrating all the force contributions: 

\begin{equation}
F_{bead}=\sum_i \frac{V_i}{V_{bead}}F_{slice}(i)
\label{eq:frac}
\end{equation}

Where $V_i$ is the volume of slice $i$ and $V_{bead}$ is the volume of the entire bead. Here we have assumed that a full calculation including Mie scattering theory is not required to obtain the qualitative dependence of trap stiffness on particle size. This assumptions is justified, since the relative refractive index between the polystyrene particle and the water surrounding it ($n_{rel}=1.6/1.33=1.2$), and the NA are small enough, no resonances are expected to occur \cite{Dunlop2008}. 
To simulate particle dynamics within the optical traps we performed Brownian dynamics simulations \cite{Volpe2013}:
\begin{equation}
x(t+\Delta t)=x(t)+\Delta t \frac{D}{k_BT}F_{bead}+\sqrt{2D\Delta t}\xi
\label{eq:brown}
\end{equation}

where $D=k_BT/\gamma$ is the diffusion coefficient, and $\xi$ is a random variable with a Gaussian distribution with zero mean and unit variance. The time interval used for the simulations was $\Delta t=10\mu s$. We adjusted rest of the simulation parameters such that the trapping stiffness for $\tilde{r}_\pi=0$ in the simulation will be equal to its value in experiment. We analyzed the simulated particle trajectories in the same manner as we did with the experimentally measured ones obtaining the simulated trap stiffness. 

The simulation result are plotted in Fig.~\ref{fig:fig5} using solid colored lines. Clearly, they are in very good agreement with the experimental measurements. 
To gain more insight into these results we simulated the dynamics of particles in a Gaussian beam and compare them to the dynamics observed in SO beams. Here we calculated the optical forces once in the Rayleigh regime (modified to account for particle size) and once using Mie scattering theory. Trap stiffness for a Gaussian beam at the whole range of $a/w$ base on Rayleigh scattering is plotted in Fig.~\ref{fig:fig5} (dashed line). The simulation result is normalized to the value of the experiment at $\tilde{r}_\pi=0$, $a=490$ nm. Even for a Gaussian beam with no side ring we observe some decrease in trap stiffness when $a\gg w$. We note that this calculation fits well only the data from the smallest particle. Next, we apply the full force calculation based on Mie scattering \cite{Nieminen2007} to a Gaussian beam (see solid black line in Fig.~\ref{fig:fig5}). We find that this calculation fits well the data from the larger particles, as long as the effect of the side ring is negligible. We note that in both cases the stiffness increases until $a>d$ and then decreases slightly. This similar qualitative behavior explains the good fit of our simplified calculation once normalized to each bead size.  

In summary, we have used SO beams to systematically measure trapping stiffness as a function of the ratio between particle diameter and the width of the beam waist in the range of $0.67<a/w<2.4$. We have found that trapping stiffness changes non-monotonically in this range and reaches a maximum around $a/w=1.5$. Interference from the side ring causes instability in trapping for larger particles. We have also provided a simplified numerical scheme to estimate the finite size effect of colloidal particles on the trapping stiffness of structured beams. This method gives very good estimates of trapping strength once calibrated against trapping in a normal beam. 

Most importantly, we have demonstrated optical trapping well below the diffraction limit. Surprisingly, we find a sharp increase in trapping stiffness once the particle diameter becomes larger the beam waist (Fig.~\ref{fig:fig5} blue spheres). This result suggests a method to enhance the trapping of nano-particles by using SO beams to create sub-diffraction-limited traps. Furthermore, SO beams can be shaped to have multiple sub-diffraction-limited features \cite{SinghSO2017,Shapira2019} in which several nano-particle can be manipulated simultaneously and in which the side ring is pushed further away from the focal point.
\newline
\newline
\noindent{ \bf Funding}. Chief scientist of Israel, Kamin project no. 55305 and Israel Science Foundation Grant no. 988/17.  
\newline


\end{document}